\documentclass[fleqn,twoside]{article}
\usepackage{espcrc2}

\newcommand{\AmS}{{\protect\the\rmfont2
  A\kern-.1667em\lower.5ex\hbox{M}\kern-.125emS}}

\title{Signatures of primordial helicity in the CMBR}

\author{L. Pogosian\address{Theoretical Physics Group,
        Imperial College, London SW7 2BZ, United Kingdom},
        T. Vachaspati\address{Department of Physics, Case Western Reserve
        University, Cleveland, OH 44106-7079,
        USA} and
        S. Winitzki\address{Ludwig Maximilians Universit\"{a}t, Sektion Physik, 80333 M\"{u}nchen,
Germany}}

\begin{document}

\begin{abstract}
We have studied ways in which helical primordial magnetic fields
could be constrained by measurements of the cosmic microwave
background radiation (CMBR). If there were helical flows in the
primordial plasma at the time of recombination, they would produce
parity violating temperature-polarization correlations ($C_l^{TB}$
and $C_l^{EB}$). However, the magnitude of helical flows induced
by helical magnetic fields is unobservably small. We discuss an
alternate scheme for extracting the helicity of a stochastically
homogeneous and isotropic primordial magnetic field using Faraday
rotation measure maps of the CMBR and the power spectrum of B-type
polarization ($C_l^{BB}$). \vspace{1pc}
\end{abstract}

% typeset front matter (including abstract)
\maketitle

\section{Introduction}

The improving quality of the CMBR measurements has made it
possible to test different models of Early Universe physics. As
more refined observations are made, a larger array of theoretical
ideas will be put to the test and more details of the history of
the universe will emerge.

A possibility that has already received some attention is that
large-scale Parity (P) violation may be observed via the CMBR
\cite{ScaFer97,LueWanKam99}. In Ref.~\cite{ScaFer97} it was shown
that a coherent magnetic field would induce non-zero P-violating
correlations in the CMBR through Faraday rotation, while in
Ref.~\cite{LueWanKam99} the P violation was due to the dynamics of
a pseudoscalar field. In Ref.~\cite{PogVacWin02} we have examined
the consequence of yet another possible source of P-violation (and
also CP-violation), namely large-scale primordial helical magnetic
fields\footnote{A vector field ${\bf v}$ is helical if $\left<{\bf
v}\cdot (\nabla \times {\bf v})\right> \ne 0$ }. Helical fields
could be expected from cosmic events such as electroweak (EW)
baryogenesis \cite{FieCar00,Cor97,Vac01a,Vac01b}. The reason for
these expectations is that within the EW model production of
baryons is accompanied by a change in the Chern-Simons number
(CS), which can be interpreted as the net helicity in the $SU(2)_L
\times U(1)_Y$ gauge fields. Changes in the CS are achieved via
the production and decay of non-perturbative field configurations,
such as sphalerons and linked loops of electroweak strings
\cite{AchVac00}. Because these configurations produce magnetic
fields, it is possible that the helicity in the non-Abelian fields
associated with the CS could be inherited by magnetic fields
present immediately after the EW phase transition. It has also
been argued in Ref.~\cite{Vac01a} that such helical magnetic
fields could remain frozen in the primordial plasma and, provided
sufficiently effective inverse cascading, develop cosmologically
interesting strengths at the time of recombination.

Our approach in this work was to assume existence of a stochastic
homogeneous and isotropic helical magnetic field at recombination
and try to answer the following three questions, treated as
independent from each other: {\bf 1)} Could helical magnetic
fields produce helical flows (kinetic helicity) in the primordial
plasma? {\bf 2)} Can one constrain primordial kinetic helicity
using the CMBR? {\bf 3)} Can one detect helical magnetic fields
using the CMBR? We provide answers in the following sections.

\section{Kinetic helicity from magnetic helicity?}
\label{kinmag}

We are interested in the effects of a statistically homogeneous
and isotropic magnetic field, with possibly non-vanishing
helicity. If we denote the Fourier amplitudes of the magnetic
field by ${\bf b}({\bf k})$, then
\begin{eqnarray}
&& \left< b_i ({\bf k} ) b_j ({\bf k}' ) \right> = (2\pi )^3
\delta^{(3)}({\bf k} + {\bf k}' )
\nonumber \\
&& \times [ (\delta_{ij} - {\hat k}_i {\hat k}_j) S(k) + i
\varepsilon_{ijl} {\hat k}_l A(k) ] . \label{bcorr}
\end{eqnarray}
Here $S(k)$ denotes the symmetric part and $A(k)$ the
antisymmetric part of the correlator.  One can check that
$\left<{\bf B} ({\bf x}) \cdot [\nabla \times {\bf B}({\bf
x})]\right>$ only depends on $A(k)$ and not on $S(k)$. Therefore
$A(k)$ represents the helical component of the magnetic field and
$S(k)$ the non-helical component.

In the tight-coupling approximation it is the Lorentz force that
drives flows in neutral plasma. An evaluation shows that the
Lorentz force depends on $S(k)$ but has no dependence on $A(k)$.
Therefore the velocity flow at last scattering is unaffected by
the helical component of the magnetic field. In reality the
coupling of photons to electrons is much stronger than that to
protons and so the plasma at recombination is better treated as
composed of two fluids: the electron-photon fluid and the proton
fluid. A calculation, presented in full in Ref.~\cite{PogVacWin02}
and closely following that of Harrison\cite{Har70}, shows that the
electron-photon fluid will gain an angular velocity ${\bf
\omega}_e = ({en_e})^{-1} \nabla^2 {\bf B}$, where $e$ and $n_e$
are the electron charge and the number density. If we estimate
$|\nabla^2 {\bf B}| \sim B/L^2$, where $L$ is the coherence scale
of the field, we find $|{\bf v}| \sim |L {\bf \omega}_e| \sim
10^{-18} \left ( {{B_0}/ {10^{-9}{\rm G}}} \right ) \left ( {{1
{\rm kpc}}/{L_0}} \right )$ where $B_0$ and $L_0$ are the magnetic
field strength and coherence scale at the present epoch. Compared
to the velocities induced by gravitational perturbations ($\sim
10^{-5}$) the velocities induced by helical fields are
insignificant.

\section{Signatures of kinetic helicity}

The CMBR anisotropies sourced by velocity flows are predominantly
due to the Doppler effect. Observations of CMBR are usually
presented in the form of spectral functions $C_l^{X,Y}$, where $X$
and $Y$ stand for $T$ (brightness temperature), $E$ or $B$
(so-called E- or B-type polarization) \cite{HuWhi97}. The
correlators $C_{l}^{TB}$ and $C_{l}^{EB}$ are parity-odd, while
all other correlators $C_{l}^{XY}$ are parity-even. The presence
of parity-violating (helical) flows will produce nonzero
$C_{l}^{TB}$ and $C_{l}^{EB}$.

In Ref.~\cite{PogVacWin02} we have shown that $C_{l}^{TB}$ and
$C_{l}^{EB}$ can be written as integral expressions depending on
the Fourier transform of the average kinetic helicity $\left<{\bf
v}\cdot (\nabla \times {\bf v})\right>$. We have assumed a power
law $k$-dependence for the relevant power spectra with a spectral
index $n$, and introduced a characteristic scale $k_*$ and a
characteristic strength $v_0$ of the helical flow. We then
evaluated $C_{l}^{TB}$ and $C_{l}^{EB}$ for several different
values of $n$ and $k_*$ to see if observations could, in
principle, constrain $v_0$. We found that for $n>-3$ the bound is
set by the cut-off provided by $k_*$. Only for smaller $n$ it is
possible to constrain $v_0$.

Causality does, in general, constrain the value of the spectral
index: $n\ge 2$. Thus, our analysis suggests that the CMBR will
not be able to constrain primordial kinetic helicity unless
helical flows were correlated on superhorizon scales.

\section{A strategy to detect magnetic helicity}
\label{strategy} From previous sections one concludes that only
the non-helical component of the magnetic field can have a
signature in the Doppler contribution to the CMBR. If we could
find another observable that is sensitive to both the non-helical
and the helical components, we could combine observations and
extract the helical component of the magnetic field. Such an
observable is the Faraday rotation of linearly polarized sources
due to light propagation through a magnetized plasma. The CMBR is
expected to be linearly polarized and so any intervening magnetic
fields will rotate the polarization vector by an angle
\begin{equation}
\theta = {3 \over {2 \pi e}} \lambda_0^2 \int \dot{\tau}({\bf x})
\ {\bf {\tilde B}} \cdot d {\bf l} \label{theta2}
\end{equation}
where $\dot{\tau}({\bf x}) \equiv n_e \sigma_T a$ is the
differential optical depth along the line of sight, $\lambda_0$ is
the observed wavelength of the radiation and ${\bf {\tilde B}}
\equiv {\bf B} a^2$ is the ``comoving'' magnetic field.

Faraday rotation depends on the free electron density, which
becomes negligible towards the end of recombination. Therefore,
the bulk of the rotation is produced during a relatively brief
period of time when the electron density is sufficiently low for
polarization to be produced and yet sufficiently high for the
Faraday rotation to occur. The average Faraday rotation (in
radians) between Thomson scatterings due to a tangled magnetic
field was calculated in Ref.~\cite{Harari} to be $\approx 0.08
\left( {B_0 \over 10^{-9}{\rm G}} \right) \left( {30{\rm GHz}
\over \nu_0} \right)^2$, where $B_0$ is the current amplitude of
the field and $\nu_0$ is the radiation frequency observed today.
The amplitude of the CMB polarization fluctuations is expected to
be an order of magnitude lower than that of the temperature
fluctuations. As discussed in Ref.~\cite{KosLo96}, detecting a
Faraday rotation of order $1^{\rm o}$ will require a measurement
which is superior in sensitivity by another factor of $10^{2}$.
Such accuracy is at the limit of current experimental proposals
but there is a hope that it will eventually be accomplished.

A polarization map of the CMBR at several wavelengths will (in
principle) make it possible to obtain a wavelength independent
``rotation measure'' (RM). The expression for RM is that of
eq.~(\ref{theta2}) divided by $\lambda_0^2$. One could then use a
``rotation measure map'' to find correlations of the RM: $RR'
\equiv \left< {\rm RM}(\hat{{\mathbf n}}) {\rm RM}(\hat{{\mathbf
n}}') \right>$, where $\hat{{\mathbf n}}$ and $\hat{{\mathbf n}}'$
are two directions on the sky. Using Eq.~(\ref{bcorr}) we find
\begin{equation}
RR' = \left({3 \over {2 \pi e}} \right)^2 \int {{d^3k}\over {(2\pi
)^3}} \left [ \alpha S(k) + \beta A(k) \right ] , \label{RR'}
\end{equation}
where $\alpha$ and $\beta$ are calculable functions given
explicitly in Ref.~\cite{PogVacWin02}. A crucial feature of $RR'$
is that it depends on both the helical and non-helical spectral
functions $S(k)$ and $A(k)$.

The polarization spectra due to Doppler effect from plasma flows
induced by tangled magnetic fields have already been calculated by
Seshadri and Subramanian \cite{SesSub01}. They computed the
correlator $C_l^{BB}$, which only depends on the non-helical
spectral function $S(k)$. This is because, as discussed in the
previous section, $A(k)$ is the force-free component of the
magnetic field and does not induce any velocity in the last
scattering surface. Hence, if we could use $C_l^{BB}$ to obtain
$S(k)$ -- which would only be possible assuming some functional
form (such as a power law) for $S(k)$ since $S(k)$ occurs within
some integrals -- we could insert the result in the expression for
$RR'$ given in Eq.~(\ref{RR'}). This will isolate $A(k)$ in
Eq.~(\ref{RR'}) and, with some assumptions about the functional
form of $A(k)$, the cosmic magnetic helicity can be evaluated.

\section{Summary}
\label{conclusions} If there is kinetic helicity at last
scattering, it would imprint a signature in the cross-correlators
$C_l^{TB}$ and $C_l^{EB}$. Kinetic helicity can be induced by
helical magnetic fields but the effect is too small to be
significant since the helical component of magnetic fields is
force-free. Instead we have proposed another strategy for
detecting the helicity of primordial magnetic fields using
polarization and rotation measure maps of the CMBR.

\end{document}